\documentstyle[aps,12pt,epsf,amssymb]{revtex}
\begin{document}
\baselineskip=0.8cm
\newcommand{\ini}{\begin{equation}}
\newcommand{\fin}{\end{equation}}
\newcommand{\inir}{\begin{eqnarray}}
\newcommand{\finr}{\end{eqnarray}}
\newcommand{\inif}{\begin{figure}}
\newcommand{\finf}{\end{figure}}
\newcommand{\bc}{\begin{center}}
\newcommand{\ec}{\end{center}}
\def\ol{\overline}
\def\pa{\partial}
\def\ra{\rightarrow}
\def\ts{\times}
\def\df{\dotfill}
\def\bs{\backslash}
\def\dg{\dagger}

$~$

\hfill DSF-06/2000

\vspace{1 cm}

\centerline{\LARGE{Seesaw mechanism, quark-lepton symmetry}}

\centerline{\LARGE{and Majorana phases}}

\vspace{1 cm}

\centerline{\large{D. Falcone}}

\vspace{1 cm}

\centerline{Dipartimento di Scienze Fisiche, Universit\`a di Napoli,}
\centerline{Mostra d'Oltremare, Pad. 19, I-80125, Napoli, Italy}

\centerline{{\tt e-mail: falcone@na.infn.it}}

\vspace{1 cm}

\begin{abstract}

\noindent
A previous short analysis of the seesaw mechanism, based on quark-lepton
symmetry, experimental data and hierarchical neutrino spectrum, is
enlarged to include small but not zero $U_{e3}$, inverted mass hierarchy,
and the qualitative effect of Majorana phases. The structure of the heavy
neutrino mass matrix obtained in several cases is discussed.
We find two leading forms for this matrix. One is diagonal
and stands at the unification scale or above. The other is off-diagonal and
stands at the intermediate scale. 



\end{abstract}

\newpage

\section{Introduction}

\noindent
The SuperKamiokande Collaboration has recently confirmed the oscillation of
atmospheric neutrinos \cite{sk}.
This evidence, as well as the strong indications of
oscillation of solar neutrinos
too, which could explain the solar neutrino deficit \cite{dhhgb,bks},
lead to nonzero neutrino masses.
Although not zero, such masses have to be much smaller than the charged lepton
and quark masses, less than few eV \cite{bww}.
This feature can be explained by means of the seesaw mechanism \cite{seesaw},
where the mass matrix $M_L$ of light (left-handed) Majorana neutrinos
is given by
\ini
M_L=M_D M_R^{-1} M_D^T,
\fin
with the Dirac mass matrix $M_D$ of the same order of magnitude of the
charged lepton or quark mass matrix, and the eigenvalues of $M_R$, the mass
matrix of right-handed neutrinos, much bigger than the elements of $M_D$.

In the Minimal Standard Model plus three right-handed neutrinos,
the mass matrix of
heavy neutrinos is generated by a Majorana mass term
$(1/2) \ol{\nu}_R M_R (\nu_R)^c$ and hence $M_R$ is not constrained.
Instead, in Grand Unified Theories (GUTs) like $SO(10)$, $M_R$ is obtained from
the Yukawa coupling of right-handed neutrinos with the Higgs field that
breaks the unification or the intermediate group to the Standard Model
\cite{moha}.
When such a field gets a VEV $v_R$, which is the unification
or the intermediate scale, the right-handed neutrinos take a mass and
$M_R=Y_R v_R$,
where $Y_R$ is the matrix of Yukawa coefficients.
Actually, this happens when and because at the same stage it is also $B-L$
broken, allowing for Majorana masses. 
In the supersymmetric case
$v_R$ is the unification scale ($v_R \sim 10^{16}$ GeV),
while in the nonsupersymmetric case it is 
the intermediate scale ($v_R \sim 10^9-10^{13}$ GeV) \cite{dk}.
On the other hand, GUTs generally predict $M_D \sim M_u$, where $M_u$ is the
mass matrix of up quarks, and $M_l \sim M_d$, where $M_l$ is the mass matrix
of charged leptons and $M_d$ the mass matrix of down quarks. This is
called quark-lepton symmetry.

From the experimental data on neutrino masses and mixings, and the quark-lepton 
symmetry, it is possible to infer the heavy neutrino mass matrix $M_R$
by inverting formula (1),
\ini
M_R=M_D^T M_L^{-1} M_D.
\fin
In fact, $M_L$ can be obtained, at least approximately, from experimental data
on neutrino oscillations, and quark-lepton symmetry suggests
\ini
M_D \simeq \frac{m_{\tau}}{m_b}~ $diag$ (m_u,m_c,m_t).
\fin
The nearly diagonal form of $M_D$ is due to the fact that mixing in the Dirac
sector is similar to the small mixing in the up quark sector \cite{rrr},
and the factor
$m_{\tau}/m_b \equiv k$ is due to approximate running from the unification or
intermediate scale, where $m_b=m_{\tau}$ should hold \cite{acpr}.
As a matter of fact $M_D$ is almost scale independent. Then the Dirac masses
of neutrinos are
fixed by the values of the up quark masses at the unification scale in the
supersymmetric model, and at the intermediate scale in the nonsupersymmetric
model. However, in both cases their values are roughly similar \cite{fk},
namely $M_D \simeq~$diag$(0.001,0.3,100)$ GeV.
It is now important to check if
the resulting scale of $M_R$ is in accordance with the physical scales of
GUTs, and also the structure of $M_R$, which would give further insight
towards a more complete theory.
This program has been addressed in refs.\cite{smirnov,llsv}
and in the recent papers \cite{af1,fal1,abr,kwm,fal2}. In this paper we want
to extend the analysis of ref.\cite{fal2}, in order to include
small but not zero
$U_{e3}$, inverted hierarchy of light neutrino masses,
approximate effect of Majorana
phases, and a discussion on the structure of $M_R$.

In section II we summarize the experimental informations on neutrino masses
and mixings, coming mainly from solar and atmospheric oscillations.
In sections III
and IV the normal and inverted mass hierarchy cases are studied. In section V
the effect of Majorana phases is briefly considered and finally we give
some concluding remarks.

\section{Neutrino masses and mixings}

\noindent
We denote by $m_i$ ($i=1,2,3$) the light neutrino masses. The mass eigenstates
$\nu_i$ are related to the weak eigenstates $\nu_{\alpha}$
($\alpha=e,\mu,\tau$) by the unitary matrix $U$,
\ini
\nu_{\alpha}=U_{\alpha i} \nu_i.
\fin
The results on solar oscillations imply for the three solutions
of the solar neutrino problem,
namely small mixing MSW (SM), large mixing MSW (LM)
and vacuum oscillations (VO),
the following orders of magnitude for $\Delta m^2_{sol}$ \cite{bks}:
\ini
\Delta m^2_{sol} \sim 10^{-6}~$eV$^2~~$(SM)$
\fin
\ini
\Delta m^2_{sol} \sim 10^{-5}~$eV$^2~~$(LM)$
\fin
\ini
\Delta m^2_{sol} \sim 10^{-10}~$eV$^2~~$(VO)$
\fin
On the other hand, atmospheric oscillations give \cite{sk}
\ini
\Delta m^2_{atm} \sim 10^{-3}~$eV$^2,
\fin
so that $\Delta m^2_{sol} \ll \Delta m^2_{atm}$. We can set \cite{af}
\ini
\Delta m^2_{sol} = m_2^2-m_1^2,~~\Delta m^2_{atm}=m_3^2-m_{1,2}^2,
\fin
and, assuming without loss of generality $m_3>0$, there are three possible
spectra for $m_i$ \cite{af}:
\ini
m_3 \gg |m_2|,|m_1|~~~$(hierarchical)$
\fin 
\ini
|m_1| \sim |m_2| \gg m_3~~~$(inverted hierarchy)$ 
\fin
\ini
|m_1| \sim |m_2| \sim  m_3~~~$(nearly degenerate).$
\fin
Moreover, due to the near maximal mixing of atmospheric neutrinos \cite{sk}
and the smallness of $U_{e3}$ \cite{chooz}, the mixing matrix $U$ can be
written as \cite{akh}
\ini
U=\left( \begin{array}{ccc}
  c & s & \epsilon \\
  -\frac{1}{\sqrt2}(s+c\epsilon) & \frac{1}{\sqrt2}(c-s\epsilon) &
   \frac{1}{\sqrt2} \\
   \frac{1}{\sqrt2}(s-c\epsilon) & -\frac{1}{\sqrt2}(c+s\epsilon) &
   \frac{1}{\sqrt2} 
    \end{array} \right),
\fin
where $\epsilon$ is small and $s=\sin\theta$, $c=\cos\theta$, with $\theta$
the mixing angle of solar neutrinos. The SM solution corresponds to
$s \simeq 0$, while the LM and especially the VO solutions correspond to
$s \simeq 1/\sqrt2$ \cite{p}, that is bimaximal mixing \cite{bim}.

We set $D_L=~$diag$ (m_1,m_2,m_3)$. Since the mixing in the charged lepton
sector can be considered small \cite{rrr} and our experimental
informations on neutrinos are
approximate, for our analysis we can also set $U^{\dg} M_L U^*=D_L$
(exact in the basis where $M_l$ is diagonal), that is
\ini
M_L=U D_L U^T,
\fin
which gives the light neutrino mass matrix \cite{akh},
valid up to small corrections of the order $\epsilon^2 \lesssim 0.03$,
\ini
M_L= \left( \begin{array}{ccc}
      \mu & \delta & \delta' \\
      \delta & \rho & \sigma \\
      \delta' & \sigma & \rho'
      \end{array} \right),
\fin
with
$$
\mu=m_1 c^2+ m_2 s^2
$$
$$
\mu'=m_1 s^2+ m_2 c^2
$$
$$
\delta=\frac{1}{\sqrt2}[\epsilon (m_3-\mu)+(m_2-m_1)cs]
$$
$$
\delta'=\frac{1}{\sqrt2}[\epsilon (m_3-\mu)-(m_2-m_1)cs]
$$
$$
\sigma= \frac{1}{2}(m_3-\mu')
$$
$$
\rho=\frac{1}{2}[ m_3+\mu'-2(m_2-m_1)cs\epsilon]
$$
$$
\rho'=\frac{1}{2}[ m_3+\mu'+2(m_2-m_1)cs\epsilon].
$$
The inverse of $M_L$ is given by
\ini
M_L^{-1}= \left( \begin{array}{ccc}
\rho \rho'-\sigma^2 & \sigma \delta'-\delta \rho' & \delta \sigma-\rho \delta'
\\ \sigma \delta'-\delta \rho' & \mu \rho'-\delta'^2 &
\delta \delta'-\mu \sigma \\ \delta \sigma-\rho \delta' &
\delta \delta'-\mu \sigma & \mu \rho-\delta^2
      \end{array} \right) \frac{1}{D},
\fin
with $D=m_1 m_2 m_3$.

In the following sections we will study the matrix $M_R$ which is obtained
from eqns.(14),(3),(2) by the first two possible neutrino spectra (10),(11) and
$s \simeq 0$ (single maximal mixing) or $s \simeq 1/\sqrt2$
(double maximal mixing). We do not consider
the nearly degenerate spectrum
because it suffers from a number of instabilities \cite{akh}. Notice
that one of the advantages of such spectrum was the possibility of providing a
hot dark matter component (with $m_i \simeq 2$ eV),
but now we believe that the amount of hot dark matter is probably
much smaller, and one neutrino with mass about 0.07 eV, as in the
hierarchical case, can be relevant \cite{gg}.
In any case, if one assumes a hierarchical $M_D$ it is quite difficult to
make $M_L$ having degenerate eigenvalues. Nevertheless, we give here a rough
evaluation for the scale of $M_R$ at the intermediate value $10^{12}$ GeV.

In this paper we do not consider the results of the LSND experiment
\cite{lsnd}, which have not yet been confirmed by other experiments. If
confirmed the LSND results would imply a third $\Delta m^2$ scale,
$\Delta m^2_{LSND} \sim 1$ eV$^2$, and thus a fourth (light and sterile)
neutrino. 

\newpage

\section{Hierarchical spectrum}

\noindent
In this case the light neutrino mass matrix can be written as
\ini
M_L= \left( \begin{array}{ccc}
      \mu & \delta & \delta' \\
      \delta & \frac{m_3}{2} & \frac{m_3}{2} \\
      \delta' & \frac{m_3}{2} & \frac{m_3}{2}
      \end{array} \right),
\fin
with
$$
\mu=m_1 c^2+ m_2 s^2
$$
$$
\delta=\frac{1}{\sqrt2}[\epsilon m_3+(m_2-m_1)cs]
$$
$$
\delta'=\frac{1}{\sqrt2}[\epsilon m_3-(m_2-m_1)cs].
$$
The leading form is
$$
M_L \sim \left( \begin{array}{ccc}
      0 & 0 & 0 \\
      0 & 1 & 1 \\
      0 & 1 & 1
      \end{array} \right).
$$
The inverse of $M_L$ is given by
\ini
M_L^{-1}\simeq \left( \begin{array}{ccc}
      m_3 \mu' & \frac{m_3}{2}(\delta'-\delta) &
      \frac{m_3}{2}(\delta-\delta') \\
     \frac{m_3}{2}(\delta'-\delta)  & \frac{m_3}{2}\mu-\delta'^2 & 
       \delta\delta'-\frac{m_3}{2}\mu  \\
     \frac{m_3}{2}(\delta-\delta')  & \delta\delta'-\frac{m_3}{2}\mu & 
     \frac{m_3}{2}\mu-\delta^2
      \end{array} \right) \frac{1}{D},
\fin
where for the entry 1-1 we have used a better degree of approximation
from eqn.(16) than that obtained from eqn.(17).
Due to the mass hierarchy (10)
we also have $m_3^2 \simeq \Delta m^2_{atm}$, for example we can take
$m_3=6 \cdot 10^{-2}$ eV. It will be useful to match results obtained
for the scale of $M_R$ with the one obtained when $M_L=D_L$, that is
$$
M_{R33} \sim \frac{k^2 m_t^2}{m_3}.
$$
Within the paper we assume that the largest Yukawa coefficient in $Y_R$
is of order 1, as indeed it happens for the up quark Yukawa coefficients.

\subsection{Single maximal mixing}

\noindent
If $s \simeq 0$, then $\delta=\delta'=(1/\sqrt2)\epsilon m_3$, so that
\ini
M_L^{-1}\simeq \left( \begin{array}{ccc}
      m_3 m_2 & 0 & 0 \\
     0  & x & -x \\
     0 & -x & x
     \end{array} \right) \frac{1}{D},
\fin   
with $x=m_3(m_1-\epsilon^2 m_3)/2$ and hence
\ini
M_{R33} \sim \frac{1}{2} \frac{m_1-\epsilon^2 m_3}{m_1 m_2} k^2 m_t^2.
\fin
If $\epsilon^2 m_3 \ll m_1$, then \cite{fal2}     
\ini
M_{R33} \sim \frac{1}{2} \frac{k^2 m_t^2}{m_2}.
\fin
Since $s \simeq 0$ corresponds to the SM solution, one has
$m_2 \lesssim 10^{-3}$ eV and $M_{R33} \gtrsim 10^{15}$ GeV.
The scale can be lowered for $m_1 \simeq \epsilon^2 m_3$.
If this cancellation does not occur, the structure of $M_R$ is hierarchical,
with the leading form
\ini
M_R \sim ~$diag$ (0,0,1),
\fin
which is the same as that obtained for $M_D$ (see eqn.(3)).

\subsection{Double maximal mixing}

\noindent
For $s \simeq 1/\sqrt2$ we have three subcases: $|m_2| \gg |m_1|$,
$m_2 \simeq m_1$ and $m_2 \simeq -m_1$.

{\bf 1.} We consider now the case with $|m_2| \gg |m_1|$, where we have
$$
\delta=\frac{1}{\sqrt2}(\epsilon m_3+m_2/2)
$$    
$$
\delta'=\frac{1}{\sqrt2}(\epsilon m_3-m_2/2)
$$   
and $\mu=m_2/2$.
If $2 \epsilon m_3\ll |m_2|$, then $\delta=m_2/2 \sqrt2=-\delta'$ and
\ini
M_L^{-1}\simeq \left( \begin{array}{ccc} 
      \frac{m_3 m_2}{2} & -\frac{m_3 m_2}{2 \sqrt2} &
      \frac{m_3 m_2}{2 \sqrt2} \\
     -\frac{m_3 m_2}{2 \sqrt2}  & \frac{m_3 m_2}{4} & -\frac{m_3 m_2}{4} \\
     \frac{m_3 m_2}{2 \sqrt2} & -\frac{m_3 m_2}{4} & \frac{m_3 m_2}{4}
     \end{array} \right) \frac{1}{D}.
\fin   
The scale of $M_R$ is given by \cite{fal2}
\ini
M_{R33} \sim \frac{1}{4} \frac{k^2 m_t^2}{m_1},
\fin
and $M_{R33}\gtrsim 10^{16}$ GeV (LM) or $M_{R33} \gtrsim 10^{18}$ GeV (VO).
If $\delta \simeq 0$ or $\delta' \simeq 0$ results are similar.
We have a hierarchical structure for $M_R$, reflecting the hierarchy of Dirac
masses. The leading form is again eqn.(22).

{\bf 2.} If $m_2 \simeq m_1$, then $\delta=\delta'=(1/\sqrt2)\epsilon m_3$ and
$\mu=m_{2}$ yielding
\ini
M_L^{-1}\simeq \left( \begin{array}{ccc}
      m_3 m_2 & 0 & 0 \\
     0  & y & -y \\
     0 & -y & y
     \end{array} \right) \frac{1}{D},
\fin   
with $y=m_3(m_{2}-\epsilon^2 m_3)/2$ and
\ini
M_{R33} \sim \frac{1}{2} \frac{m_{2}-\epsilon^2 m_3}{m_{2}^2} k^2m_t^2.
\fin
If $\epsilon^2 m_3 \ll m_{2}$, then \cite{fal2}
\ini
M_{R33} \sim \frac{1}{2} \frac{k^2 m_t^2}{m_{2}}  
\fin
and $M_{R33} \gtrsim 10^{15}$ GeV (LM and VO). The scale can be lowered if
$m_{2} \simeq \epsilon^2 m_3$. If the cancellation does not occur, $M_R$ is
hierarchical with the leading form (22).

{\bf 3.} For $m_2 \simeq -m_1$ we have
$$
\delta=\frac{1}{\sqrt2}(\epsilon m_3+m_2)
$$    
$$
\delta'=\frac{1}{\sqrt2}(\epsilon m_3-m_2)
$$    
and $\mu \simeq 0$. 
Assuming $\epsilon m_3 \ll |m_2|$, one has $\delta=m_2/\sqrt2=-\delta'$ and
\ini
M_L^{-1}\simeq \left( \begin{array}{ccc}
      0 & -{\sqrt2}m_3 m_2 & {\sqrt2}m_3 m_2 \\
    -{\sqrt2}m_3 m_2  & -m_2^2/2 & -m_2^2/2 \\
     {\sqrt2}m_3 m_2 & -m_2^2/2 & -m_2^2/2
     \end{array} \right) \frac{1}{D},
\fin   
\ini
M_{R33} \sim \frac{1}{2} \frac{k^2 m_t^2}{m_3}
\fin
\ini
M_{R13} \sim \sqrt2 \frac{k^2 m_u m_t}{m_2}.
\fin
For $m_2/m_3 \sim m_u/m_t$, $M_{R33}$ and $M_{R13}$ are similar and near the
unification scale. Otherwise $M_R$ is hierarchical.
An interesting case is $\delta \simeq 0$, which is possible if $m_2 <0$, when
$\delta'=-\sqrt2 m_2$ and
\ini
M_L^{-1}\simeq \left( \begin{array}{ccc}
         0 & -\frac{m_3 m_2}{\sqrt2} & \frac{m_3 m_2}{\sqrt2} \\
         -\frac{m_3 m_2}{\sqrt2} & 2 \epsilon m_3 m_2 & 0 \\
         \frac{m_3 m_2}{\sqrt2} & 0 & 0
\end{array} \right) \frac{1}{D},
\fin
so that the scale is given by
\ini
M_{R13} \sim \frac{1}{\sqrt2} \frac{k^2 m_u m_t}{m_{2}}
\fin
that is intermediate. In fact $m_2 \lesssim 10^{-3}$ eV gives
$M_{R33} \gtrsim 10^{11}$ GeV.
In this special case the structure of $M_R$ is
roughly off-diagonal with the leading form
\ini
M_R \sim \left( \begin{array}{ccc}
     0 & 0 & 1 \\ 0 & 0 & 0 \\ 1 & 0 & 0
     \end{array} \right),
\fin
which was obtained for example in refs.\cite{js,bs}. 
If $\delta' \simeq 0$, $\delta=\sqrt2 m_2$ and 
\ini
M_L^{-1}\simeq \left( \begin{array}{ccc}
         0 & -\frac{m_3 m_2}{\sqrt2} & \frac{m_3 m_2}{\sqrt2} \\
         -\frac{m_3 m_2}{\sqrt2} & 0 & 0 \\
         \frac{m_3 m_2}{\sqrt2} & 0 & 2 \epsilon m_3 m_2
\end{array} \right) \frac{1}{D},
\fin
with $M_{R33} \sim \epsilon k^2 m_t^2/m_2$, $M_{R13} \sim k^2 m_u m_t/m_2$,
near the unification scale.

\section{Inverted hierarchy}

\noindent
In this case the light neutrino mass matrix is
\ini
M_L= \left( \begin{array}{ccc}
      \mu & \delta & \delta' \\
      \delta & \frac{\mu'}{2} & -\frac{\mu'}{2} \\
      \delta' & -\frac{\mu'}{2} & \frac{\mu'}{2}
      \end{array} \right),
\fin
with
$$
\mu=m_1 c^2 +m_2 s^2
$$
$$
\mu'=m_1 s^2 +m_2 c^2
$$
$$
\delta=-\frac{1}{\sqrt2}[\epsilon \mu -(m_2-m_1)cs]
$$    
$$
\delta'=-\frac{1}{\sqrt2}[\epsilon \mu +(m_2-m_1)cs]
$$    
\ini
M_L^{-1}\simeq \left( \begin{array}{ccc}
    m_3 \mu' & -(\delta+\delta')\frac{\mu'}{2} & -(\delta+\delta')\frac{\mu'}{2} \\
    -(\delta+\delta')\frac{\mu'}{2} & \frac{\mu \mu'}{2}-\delta'^2 &
    \frac{\mu \mu'}{2}+\delta \delta' \\
    -(\delta+\delta')\frac{\mu'}{2} & \frac{\mu \mu'}{2}+\delta \delta' &
     \frac{\mu \mu'}{2}-\delta^2
  \end{array} \right) \frac{1}{D},
\fin
and $m_{1,2}^2 \simeq \Delta m^2_{atm}$.
The lightest neutrino mass $m_3$ does not depend on the solar
neutrino solution, and can be arbitrarily small.

\subsection{Single maximal mixing}

\noindent
If $s \simeq 0$, then $\mu=m_1$, $\mu'=m_2$,
$\delta=-(1/\sqrt2)\epsilon m_1=\delta'$,
the leading $M_L$ is given by
$$
M_L \sim \left( \begin{array}{ccc}
      1 & 0 & 0 \\
      0 & 1 & 1 \\
      0 & 1 & 1
      \end{array} \right)
$$
and the inverse of $M_L$ by
\ini
M_L^{-1}\simeq \left( \begin{array}{ccc}
        m_3 m_2 &  \epsilon \frac{m_1 m_2}{\sqrt2} & 
       \epsilon \frac{m_1 m_2}{\sqrt2}   \\
         \epsilon \frac{m_1 m_2}{\sqrt2} & \frac{m_1 m_2}{2} &
        \frac{m_1 m_2}{2} \\
        \epsilon \frac{m_1 m_2}{\sqrt2} & \frac{m_1 m_2}{2}  &
        \frac{m_1 m_2}{2}   
        \end{array} \right) \frac{1}{D}
\fin
so that
\ini
M_{R33} \sim \frac{1}{2} \frac{k^2 m_t^2}{m_3},
\fin
which is at or above the unification scale. The structure of $M_R$ is
hierarchical, with the leading form (22).

\subsection{Double maximal mixing}

\noindent
For $s \simeq 1/\sqrt2$ we have two cases, corresponding to $m_2 \simeq m_1$
and $m_2 \simeq -m_1$.

If $m_2 \simeq m_1$, then $\mu=m_{1,2}=\mu'$, 
$\delta=-(1/\sqrt2)\epsilon m_{1,2}=\delta'$,
the leading $M_L$ is like for $s \simeq 0$ and                              
\ini
M_L^{-1}\simeq \left( \begin{array}{ccc}
       m_3 m_{1,2} & \epsilon \frac{m_{1,2}^2}{\sqrt2} & 
       \epsilon \frac{m_{1,2}^2}{\sqrt2}   \\
        \epsilon \frac{m_{1,2}^2}{\sqrt2} & \frac{m_{1,2}^2}{2} &
        \frac{m_{1,2}^2}{2} \\
        \epsilon \frac{m_{1,2}^2}{\sqrt2} & \frac{m_{1,2}^2}{2} &
         \frac{m_{1,2}^2}{2}   
         \end{array} \right) \frac{1}{D} 
\fin
with the same result as for $s \simeq 0$.

If $m_2 \simeq -m_1$, then $\mu=\mu'=0$, $\delta=(1/\sqrt2) m_{1,2}=-\delta'$,
the leading $M_L$ is
$$
M_L \sim \left( \begin{array}{ccc}
      0 & 1 & 1 \\
      1 & 0 & 0 \\
      1 & 0 & 0
      \end{array} \right)
$$
and the inverse is
\ini
M_L^{-1}\simeq \left( \begin{array}{ccc}
          0 & 0 & 0 \\
          0 & -\frac{m_{1,2}^2}{2} & -\frac{m_{1,2}^2}{2} \\
          0 & -\frac{m_{1,2}^2}{2} & -\frac{m_{1,2}^2}{2} 
           \end{array} \right) \frac{1}{D} 
\fin
with
\ini
M_{R33} \sim 2 \frac{k^2 m_t^2}{m_3},
\fin
similar to above, and with a hierarchical $M_R$, of the leading form (22).

\section{Effect of phases}

\noindent
In the preceding sections we have considered only real matrices, which is a CP
conserving framework. The signs of $m_i$ correspond to CP parities of
neutrinos, while the physical masses are $|m_i|$ \cite{bgg}. Let us
now write a more general form of $M_L$ \cite{gag},
namely the same as eqn.(14) but with $U$ parametrized as the ordinary CKM
matrix (with the CP violating phase $\delta$) and
\ini
D_L=~$diag$ (m_1 e^{i \alpha}, m_2 e^{i \beta},m_3)
\fin
with $m_i>0$ \cite{bgg}. We see that the preceding formalism trasforms according
to
$$
m_1 \ra m_1 e^{i \alpha},~m_2 \ra m_2 e^{i \beta},~
\epsilon \ra \epsilon e^{i \delta}.
$$
Moreover, in the hierarchical case $\epsilon$ (or $\epsilon^2$) is often
joined to $m_3$, while in the inverted hierarchy case it is joined to
$m_{1,2}$. It is clear that if there is no fine tuning of
the parameters $m_i$, $\epsilon$,
phases have a minor effect. However, we have found some important cases
where cancellations occur, indicating also small (that is about 0) or large
(that is about $\pi$) phase differences. For example eqn.(31) may be obtained
for $\alpha \simeq 0$, $\delta \simeq 0$, $\beta \simeq \pi$. It is to remember
that only the phase $\delta$ affects neutrino oscillations (see $\epsilon$ in
eqn.(13)), while all three phases appear in the neutrinoless double-beta
decay parameter $M_{ee}=|U_{ei}^2 m_i|$. If $\epsilon \simeq 0$ and
$|m_2| \simeq |m_1|$, a large phase difference $\alpha-\beta \simeq \pi$ gives a
much smaller $M_{ee}$ with respect to a small phase difference
$\alpha-\beta \simeq 0$.

\section{Concluding remarks}

\noindent
We have found two leading forms for $M_R$, namely eqn.(22) ($M_R$ diagonal)
and eqn.(33) ($M_R$ off-diagonal). The diagonal form generally is around the
unification scale (except for the case of VO with full hierarchy, where the
scale goes well above the unification scale, towards the Planck scale
\cite{fal1}),
while the off-diagonal form is at the intermediate scale and hence welcome in
the nonsupersymmetric model. Moreover, the off-diagonal form is obtained for a
particular pattern for the signs of the light neutrino masses, namely
$m_2$ opposite to both $m_1$ and $m_3$,
with nearly bimaximal mixing.
Of course, this pattern gives a smaller $M_{ee}$ with respect
to the pattern with all $m_i$ of the same sign.

From the point of view of effective parameters, the off-diagonal form seems
related to some suitable cancellations, but all of them lead to the smallness
of entry $M_{R33}$ and hence point towards a different underlying theory with
respect to the diagonal form, where the largest element is just $M_{R33}$.
With regard to this, we would like to refer, for example, to the model \cite{bs},
where a suitable pattern of horizontal $U(1)$ charges gives
$$
M_R \simeq \left( \begin{array}{ccc}
        0 & \sigma^2 & 1 \\
        \sigma^2 & \sigma^2 & 0 \\
         1 & 0 & 0 
      \end{array} \right) M_0,
$$
with $\sigma =(m_c/m_t)^{1/2}$ and $M_0 \sim 10^{12}$ GeV.
 
$~$

$~$

The author thanks F. Buccella for discussions.

\end{document}